\documentclass[a4paper]{article}
\usepackage{amsmath}
\usepackage{graphicx}

\begin{document}
\begin{center}
\Large
\textbf{Holographic interferometry for the study of liquids}
\normalsize
\vspace{.5cm}

Jean Colombani\footnote{Corresponding author, tel.: +33 472448570,
fax.: +33 472432925, Jean.Colombani@lpmcn.univ-lyon1.fr}
and Jacques Bert
\vspace{.5cm}

Laboratoire de Physique de la Mati\`ere Condens\'ee et Nanostructures;

Universit\'e Claude Bernard Lyon 1; CNRS, UMR 5586

Domaine scientifique de la Doua, F-69622 Villeurbanne cedex; France
\end{center}

\begin{center}
\section*{Abstract}
\end{center}

Holography is an optical technique enabling to record phase objects.
Holographic interferometry uses this faculty to make a phase object
interfere with a memory of itself at a preceding time, recorded on a
hologram.
Interference fringes therefore inform on any variation of the phase
of the object.
For the study of liquids, these phase changes can result from the evolution
of temperature or concentration (via the index of refraction).
This access to the real-time evolution of concentration can be used to
measure diffusion coefficients, Soret coefficients or dissolution coefficients.
Temperature fringes can be used to study convectives flows.

\vspace{.5cm}

\textbf{Keywords:} holographic interferometry, liquids, diffusion.

\vspace{1cm}

\section{Introduction}

The expression "holographic interferometry" has a double etymologic origin:
"holography" derives from the greek words $o \lambda o\varsigma$
and $\gamma\rho\alpha\phi\epsilon\iota\nu$, meaning "to write all",
whereas the term "interferometry" comes from the latin words \textit{inter},
\textit{ferire} and \textit{metrum},
meaning literally "a measure of the hits inbetween".
We are going to try, along this article, to give a more explicit definition of
this powerful optical technique.

Holography was invented in 1948 by Dennis Gabor (1900-1979), a hungarian-british
physicist, just before its arrival at the Imperial College in London.
His first motivation was the reconstruction of wavefront in
electron microscopy \cite{Gabor48}.
He received the Nobel prize in physics in 1971 "for his invention
and development of the holographic method".

But the burst of the possible applications of holography appeared in optics.
Leith and Upatnieks, from the Radar Laboratory of the University of
Michigan, benefitted from the recently developped laser technology to
achieve the reconstruction of light wavefront, thus enabling three-dimensional
photography in 1962 \cite{Leith}.

In the following, we will first describe briefly the basic concepts of
holography and holographic interferometry and then focus on known applications
of this technique to the study of transparent liquids properties.

\section{Principle of holographic interferometry}
\label{principle}
To choose a model of light waves relevant for the description of holography,
the following approximations are made.
The frequency of light ($\sim 10^{15}$ Hz) is not accessible to
classical detector, which measure time-averaged values,
so a timeless model is enough.
Monochromatic light will be needed, so a one-wavelength model is
considered.
Linear polarisation is preferable,
so a scalar representation is sufficient.
Therefore to understand the basic principle of holography, light can be
viewed as a
mere complex amplitude $U=A\exp(i\phi)$, with $A$ the real amplitude and
$\phi$ the phase.
This phase is linked to the optical pathlength $\delta$ via
$\phi=2\pi\delta/\lambda$.
The optical pathlength contains information 
on the transmitting medium : $\delta=n\times e$
($e$ the depth and $n$ the index of refraction).

Classical optical detectors like eyes or cameras are solely sensible to
the light intensity $I=\epsilon v |U|^2\sim |U|^2 = A^2$
($\epsilon$ the permittivity  and $v$ the wave velocity).
So the phase cannot be registered and the information it contains is lost
by these sensors.
Therefore, indices of refraction and 3D vision are not accessible directly
with photography, video, \dots

To remedy this lack, Gabor has appealed to Young's theory of interference.
The main idea of this theory published in 1801 lies in the fact that when
two beams superimpose in one point, their complex amplitude must be summed up,
not their intensity, as thought before.
This explains that a combination of lights can result in darkness.
Therefore, in the case of two beams of complex amplitudes $U_1=A_1\exp(i\phi_1)$
and $U_2=A_2\exp(i\phi_2)$ crossing in one point, the resulting intensity
is $I\sim|U_1+U_2|^2 = A_1^2+A_2^2+2A_1A_2\cos(\phi_1-\phi_2)$.
One notices that this expression includes a $\phi_1-\phi_2$ term,
thus containing information on the phases.
So if the phase of one of the two beams is known, the other can be deduced.
Here lies the basic principle of holography : 
the detector is illuminated by the beam coming from the object PLUS by a
reference beam with a known phase.
Thereby the recorded interference pattern contains the phase difference
$\phi_1-\phi_2$
between the two beams, and so informations on the index of refraction and 3D 
topography of the object \cite{Vest}.

Classical holography is performed in two steps :
\begin{description}
\item[\underline{A recording step}]
A photographic plate is illuminated by the object beam and a reference beam
(generally a plane or spherical wave) of a coherent light.
The interference pattern between the two beams is recorded through grey levels
on the plate and is called the hologram.
This pattern contains a 'coding' of the object phase.
\item[\underline{A reconstruction step}]
The object is removed and the plate is developped and illuminated
by the reference beam only.
The diffraction of the reference beam by the interference pattern on the plate
provides a 'decoding' of the recorded object phase.
Therefore, when observing through the hologram, a 3D picture of the object
can be viewed.
A hologram should be viewed as a hole in a wall behind which the object stands
rather than as a photograph.
For instance, if the plate is broken in two parts, the whole object can
still be observed with each part of the hologram alone but with different
angles of view, like when hiding half of a hole in a wall.
\end{description}

\begin{figure}
\begin{center}
\includegraphics[width=.5\textwidth,angle=-90]{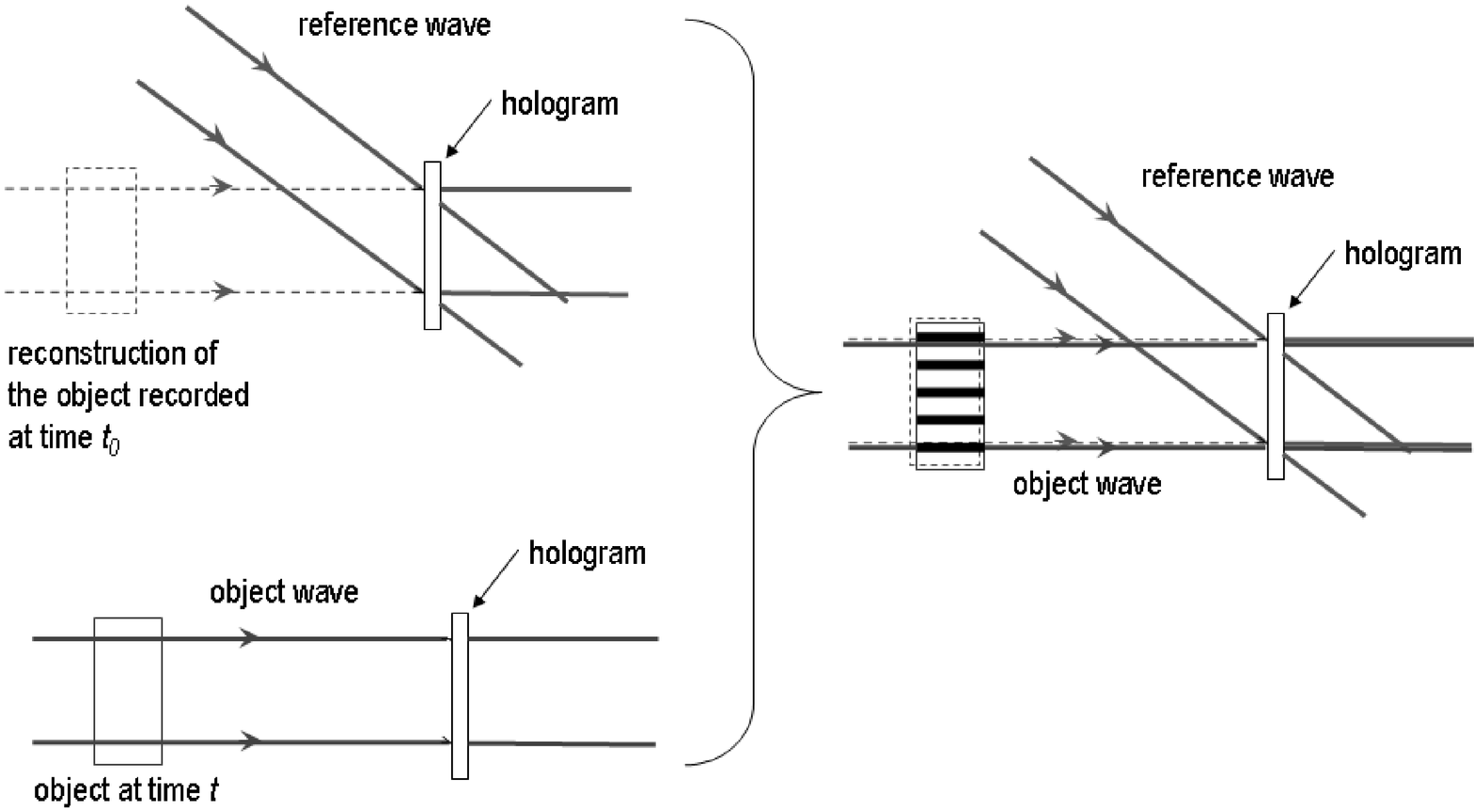}
\caption[Schematics of holographic interferometry principle.]{}
\label{interfholo}
\end{center}
\end{figure}

Holographic interferometry uses the opportunity of recording phase objects
to make an object interfere with a memory of itself at a preceding time
(registered in the hologram).
According to Vest \cite{Vest}, the application of holography to interferometry
was first suggested by Horman \cite{Horman} and Gabor et al.
\cite{Gabor65}
and developped simultaneously in several laboratories during the year 1965.
The first application of this technique to fluids was the visualization of
gas flows \cite{Brooks}.
The experiment proceeds as follows :
\begin{itemize}
\item At a reference time $t_0$, the investigated object is illuminated
with the object and reference beams and the hologram is recorded.
\item Subsequently both the reference beam ---in order to obtain the
reconstruction of the object wave at $t_0$--- and object beam ---creating
the object wave at time $t$--- are switched on briefly (double exposure
holographic interferometry) or continueously (real-time holographic
interferometry).
At the object position, one can then virtually find both the illuminated object
at present time and its 3D memory at $t_0$, reconstructed by the diffraction
of the reference beam by the hologram.
If a change in the optical pathlength has occured, interference fringes appear
(cf. Figure \ref{interfholo}).
This optical pathlength evolution may result from a motion of the object
or from a modification of its index of refraction due to a change in a
thermodynamic parameter.
So a real-time map of the temperature, concentration, pressure, stress,\dots
can be deduced from the fringe pattern \cite{Vest}.
\end{itemize}

This technique is less experimentally demanding than
classical interferometry
because all optical defects between the laser and the hologram (except in the
optical cell) remain unchanged at time $t_0$ and $t$, and therefore compensate.
Furthermore the data analysis is straightforward.

But it should be mentionned that, like for all interferometry measurements,
the initial conditions ---recorded in the reference hologram--- must be reliably
known.
Indeed interferograms give only informations on the evolution of the system
since these initial conditions.

Nowadays, classical holographic interferometry is progressively supplanted
by digital holographic interferometry \cite{Goodman}.
In this alternative technique, the holographic plate is replaced by a
Charge-Coupled Device (CCD) camera.
During the recording step the interference pattern of the object and
reference beams is recorded through grey levels by the CCD sensor.
During the reconstruction step, the diffraction of the reference beam by
the recorded interference pattern is carried out numerically instead
of physically.
This process has numerous advantages: no need of the time-consuming step
of photographic plate development, easy change of the initial time $t_0$,
straightforward access to the object phase \dots
But its drawbacks are not to be underestimated: small size of the sensor
chips compared to photographic plates (so smaller investigated objects),
small resolution
of the sensor compared to photographic plates, subtle calculations, \dots

Our experimental setup, used in all the below-mentionned studies, has been
chosen for its simplicity (no use of goniometer, auxiliary fringes, reference
zone in the hologram, \dots) and is strictly the same for each application:
\begin{itemize}
\item a green 100 mW laser;
\item two equivalent beams containing a microscope objective/pinhole/convergent
lens set expanding and filtering the beams;
\item the object inserted in one of the two beams;
\item a hologram at the interceipt of the two beams;
\item a CCD camera behind the hologram, recording the interferograms.
\end{itemize}

As a summary, holographic interferometry enables to perform interference
between objects separated in time
instead of interference between objects separated in space with
classical interferometry.
We will now turn to the experiments performed in our university, which cover the
most encountered applications of this technique to the study of liquids.

One of our purposes in presenting these examples
is to demonstrate the versatility of a simple holographic
interferometry device, enabling to perform diffusion, dissolution, convection,~
\dots studies without any change in the optical apparatus.
Another purpose is to stress on the quantitative measurements accessible in all
the cases listed below, whereas holographic interferometry is often used to get
a qualitative picture of the transfers in fluids.

\section{Possible applications in liquids}

\subsection{Diffusion}

Diffusion is perhaps the most investigated phenomenon in liquids with
holographic
interferometry and was used in physicochemistry, biology, geology, \dots.
Historically, the first liquid diffusion studies were carried out
a few years only after the invention of the technique
\cite{Becsey71,Budziak,Durou72,Durou73}.
Soon very simple versions of the holointerferometric device dedicated to
liquid diffusion appeared \cite{Bochner}, still in use nowadays \cite{Anand}.
Our setup is similar to this simplified device.

We present here two possible procedures to access to the diffusion
coefficient---
one in a homogeneous protein solution and one in an immiscible mixture---
representative of the possibilities of the technique.
We stress that we are concerned in this article with dynamic phenomena
in liquids but similar measurements are also performed in gases \cite{Baranski}
and gas-liquid systems \cite{Roetzel}.

\subsubsection{Interdiffusion}
\label{diffusion}

To measure the diffusion coefficient $D$ of an aqueous protein solution at a
given concentration $c_0$, the following process has been used.
First, half of an optical cell has been filled with a solution at a slightly
lower concentration $c_0-\delta c/2$.
Then the reference hologram has been recorded.
Subsequently, the remainder of the cell was filled with a solution at a slightly
larger concentration $c_0+\delta c/2$.
The two mixtures interdiffuse and their two concentrations progressively change.
This evolution induces a refraction index modification, so interference
fringes appear.
Two adjacent fringes show a $2\pi$ phase evolution, corresponding to
a concentration evolution
$\delta c=\lambda/(e(\partial n/\partial c))$,
where $\lambda$ is the laser wavelength, $e$ the optical pathlength in the cell
and $\partial n/\partial c$ the refraction index evolution with
concentration of the solution.
Starting from the ends of the cell (where concentration is still unchanged),
the concentration profile can then be gradually computed from fringe
to fringe.

Considering that $D$ remains constant in the small investigated concentration
range $\delta c$,
the solution of Fick's law of diffusion in this geometry gives the following
expression of the concentration evolution with space and time:
\begin{equation}
c(z,t)=c_0+\frac{\delta c}{2}\text{erfc}\frac{z}{2\sqrt{Dt}}
\label{eqdif}
\end{equation}
where $z$ is the vertical coordinate and $z=0$ the interface between
the two solutions.
A fit of the experimental concentration curves at all times with this
formula gives the value of $D$ at $c_0$ \cite{Reyes}.
This procedure was inspired by diffusion experiments by holointerferometry
in an aqueous electrolyte \cite{Aouizerat}.

\subsubsection{Diffusion through a meniscus}

\begin{figure}
\begin{center}
\includegraphics[width=.6\textwidth,angle=-90]{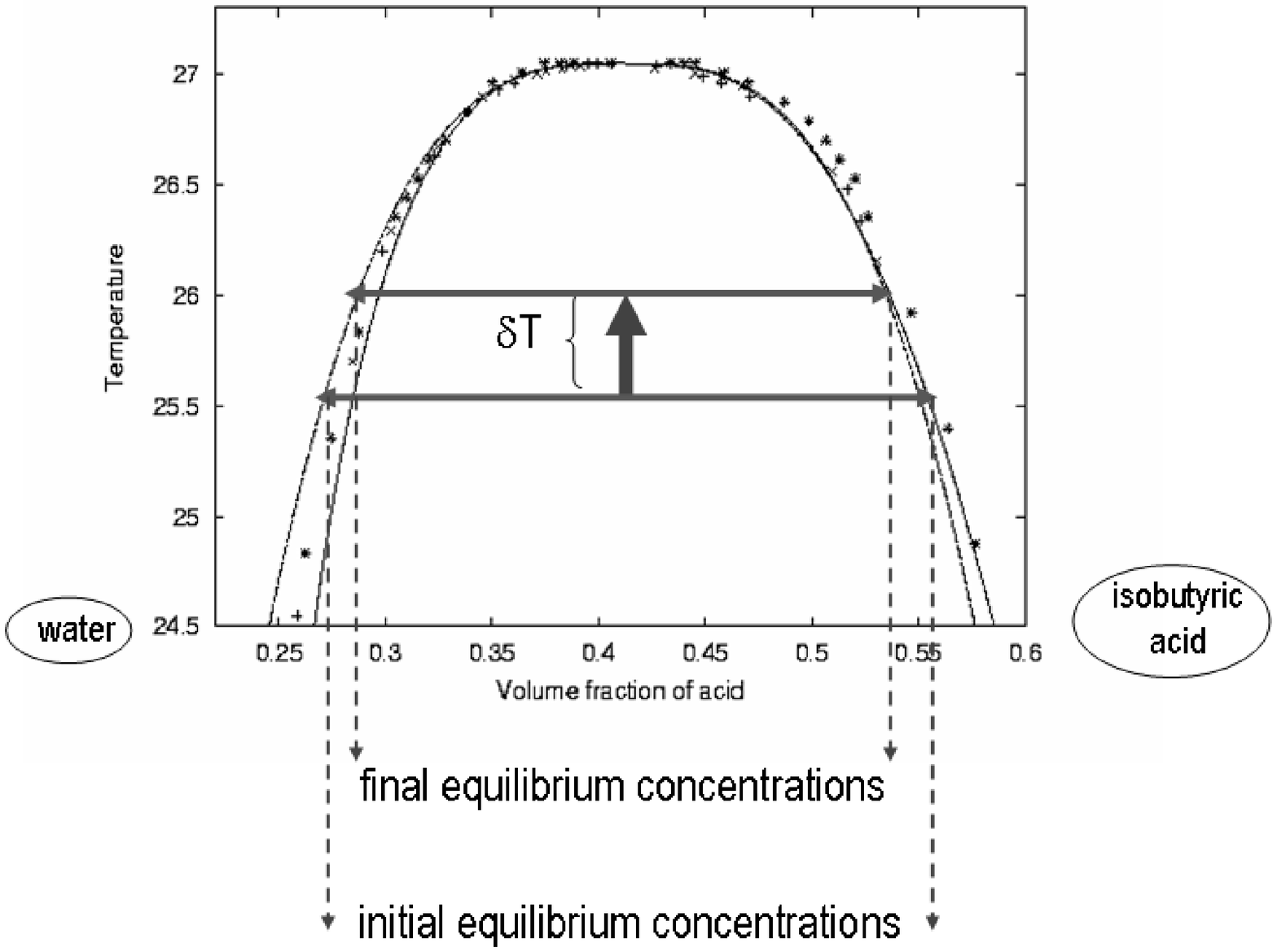}
\caption[Phase diagram of the water/isobutyric acid mixture.
The initial concentrations of the two phases and the final concentrations
after a temperature increase are also shown.]{}
\label{diagphase}
\end{center}
\end{figure}

\begin{figure}
\begin{center}
\includegraphics[width=.6\textwidth]{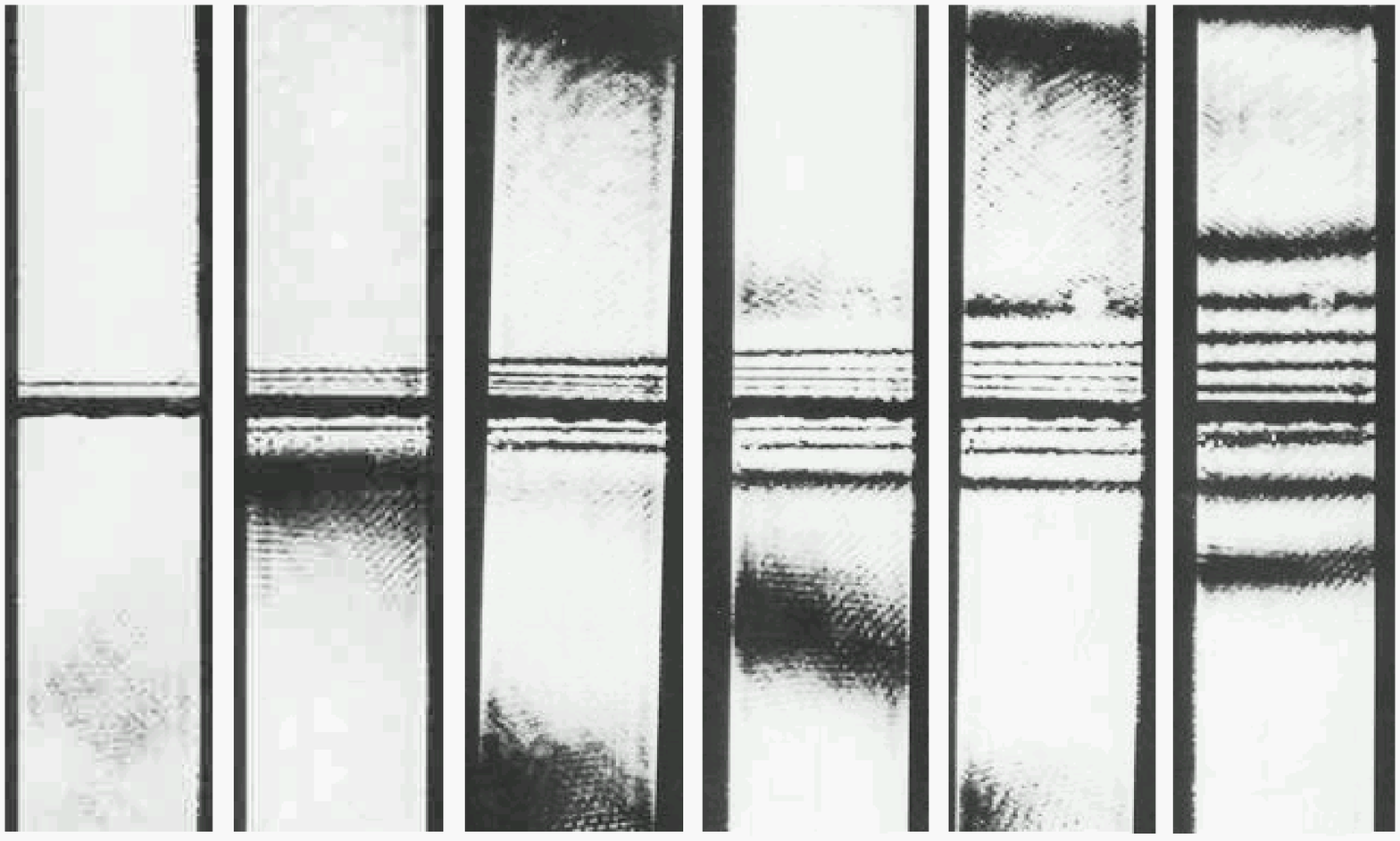}
\caption[Concentration fringes in a critical water/isobutyric acid mixture
due to the reequilibration of the concentration
in the two phases via diffusion across the meniscus
15, 90, 180, 330, 570 and 3000 minutes after an abrupt increase
of the temperature from 24.7 to 24.9$^{\circ}$C.]{}
\label{eauacide}
\end{center}
\end{figure}

In this case, our goal was to obtain values of the diffusion coefficient of the
two phases in an immiscible binary mixture, namely water/isobutyric acid.
This system exhibits a miscibility gap with an upper critical point at
$T_c=27.05^{\circ}$C.
The reference hologram is taken at a temperature below $T_c$, so in the
inhomogeneous regime, with the lighter phase standing above the heavier one.
Then temperature is abruptly increased by a few tenth of degrees.
The equilibrium concentration of the two phases therefore change
(see phase diagram in Figure \ref{diagphase}).
The new equilibrium concentrations are reached via the diffusion of water
and acid through the meniscus.
Fringes accordingly move from the meniscus toward the ends of the cell
until the two phases become homogeneous again as shown in Figure \ref{eauacide}.
As above, the real-time concentration profile $c(z)$
can be computed thanks to the knowledge
of the concentration difference between two fringes 
$\delta c=\lambda/(e(\partial n/\partial c))$.
Again the fit of the experimental concentration curves with a theoretical
law of the kind of Equation \ref{eqdif} brings
the diffusion coefficient of the two phases at the final temperature
\cite{Colombani01}.

\subsection{Soret effect}

This effect, also called thermal diffusion, is a second-order transport
phenomenon, discovered by Charles Soret in 1879: when a homogeneous
binary mixture of concentration $c$
is submitted to a thermal gradient $\nabla T$, a heat flux arises
(direct transport, Fourier's law) PLUS a matter flux tending to separate the two
components \cite{Soret}.
During the course of a Soret experiment, a concentration $\nabla c$ builds in,
leading to an opposite matter flux due to Fick's diffusion.
When the Fick and Soret flows exactly compensate, a steady state is reached
characterized by the Soret coefficient $S_T=\frac{1}{c(1-c)}
\frac{\nabla c_\infty}{\nabla T}=\frac{D'}{D}$
($D'$ thermal diffusion coefficient and $D$ diffusion coefficient).

Soret experiments are made delicate by the existence of the thermal gradient,
which is liable to induce convective disturbing currents.
Holographic interferometry provides a mean to simultaneously measure $S_T$
and check the absence of convection.
But unlike diffusion, studies of the Soret effect by
holointerferometry are rare and we were the first ones to carry out such
measurements since the pioneering works of Becsey et al.
\cite{Becsey70}
and Sanchez et al. \cite{Sanchez73,Sanchez76}.
Our device is a simplified version of theirs, without the use of
auxiliary fringes.

\begin{figure}
\begin{center}
\includegraphics[width=.6\textwidth]{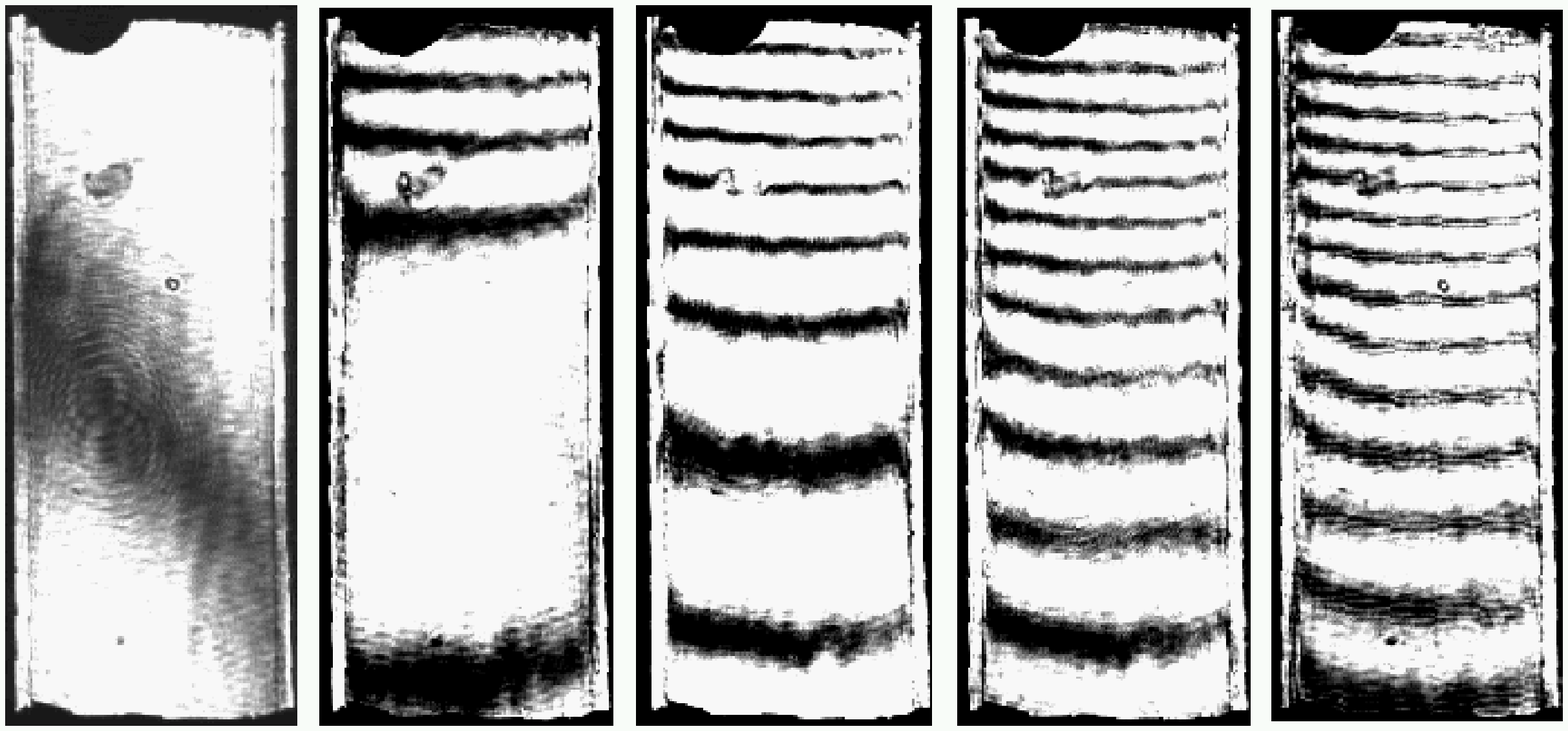}
\caption[Concentration fringes due to the Soret
effect in (LiCl,H$_2$O) 2, 48, 100, 170 and 241 minutes after applying
a 6$^{\circ}$C temperature difference between the top and bottom of the cell.
The total vertical concentration difference in the cell at the end of the
experiment is 0.5 \% in mass, corresponding to $S_T$= -3.78$\times 10^{-3}$
K$^{-1}$.]{}
\label{Soret}
\end{center}
\end{figure}

During an experiment, a hologram of an homogeneous sample of the investigated
liquid mixture is first recorded.
Then a vertical thermal gradient is imposed, leading to temperature fringes
quickly appearing.
Subsequently, concentration fringes slowly settle
due to the Soret effect.
The total concentration difference between the top and bottom of the
cell can be computed with
$\Delta c=N_c\lambda/(e(\partial n/\partial c))$
($N_c$ total number of concentration fringes in the cell).
Therewith, the evolution of the concentration gradient between the two ends
of the cell with time can be drawn, $\nabla c_\infty$ estimated and
$S_T$ calculated.
Figure \ref{Soret} shows an example of the fringe pattern evolution
during a Soret experiment ($\Delta T= 6^{\circ}$C) in a concentrated aqueous
solution of lithium chloride \cite{Colombani98,Colombani99}.

\subsection{Convection}

The three above-mentionned studies have benefitted from the holographic
concentration cartography to access to quantitative transport coefficients
in the liquid.
Besides one can also make use of the two-dimensional concentration or
temperature field to gain information about the flows in the liquid.
This has been done since the beginning of the study of liquids with
holointerferometry, the phase defect cancellation of this technique
(see Section \ref{principle}) permitting
to investigate easily large samples \cite{Vest}.
Numerous studies of this kind are still performed nowadays,
looking at heat or matter nondiffusive fluxes.

For instance, at the end of a Soret experiment, the liquid mixture exhibits
a thermal and
a solutal gradient, which result in a vertical density gradient.
From a hydrodynamic point of view, if the resulting density distribution shows
a smaller density at the top than at the bottom of the cell, the liquid
remains always stable and heat transfer proceeds by diffusion.
In the opposite case (smaller density at the bottom), for slight gradients
the liquid remains quiet, by above a threshold, an instability develops and
convection arises in the cell.

In the case of the experiment mentioned above, the solution was heated from
below, which has an unstabilizing influence,
but the Soret coefficient was negative, indicating that the salt
migrates toward the hot end of the cell, i.e., toward the cell bottom,
which has a stabilizing influence.
The hydrodynamic behaviour of the mixture in the applied temperature and
concentration fields can be summarized in a stability diagram.

\begin{figure}
\begin{center}
\includegraphics[width=4cm]{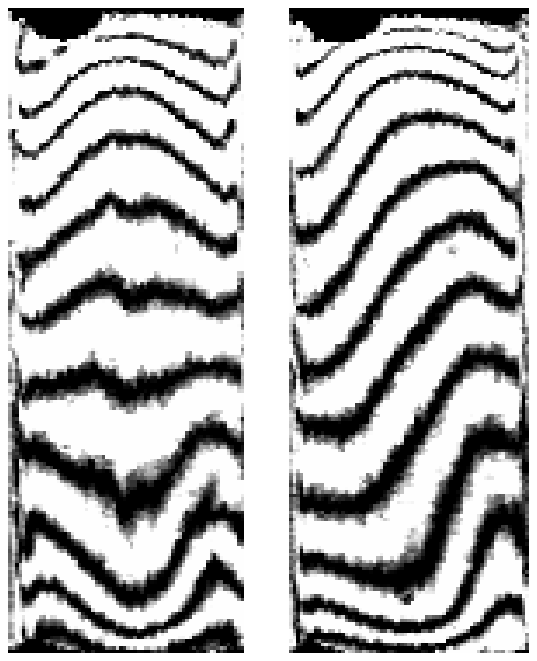}
\caption[Temperature fringes in (LiCl,H$_2$0) in the stationary convective
regime for a 10.6$^{\circ}$C temperature difference between top and bottom:
on the left, the convection roll has its axis horizontal in the plane of the
paper, on the right its axis is horizontal and perpendicular to the paper.]{}
\label{rolls}
\end{center}
\end{figure}

In a holographic interferometry experiment, the gradients are visualized
through fringes.
In a diffusive state, isotherms (so temperature fringes) stay perpendicular
to the thermal gradient (so horizontal).
In a convective state, a roll of flowing liquid develops, leading to a
distorsion of the isotherms (so of the temperature fringes), as shown in
Figure \ref{rolls}.

\begin{figure}
\begin{center}
\includegraphics[width=.6\textwidth,angle=-90]{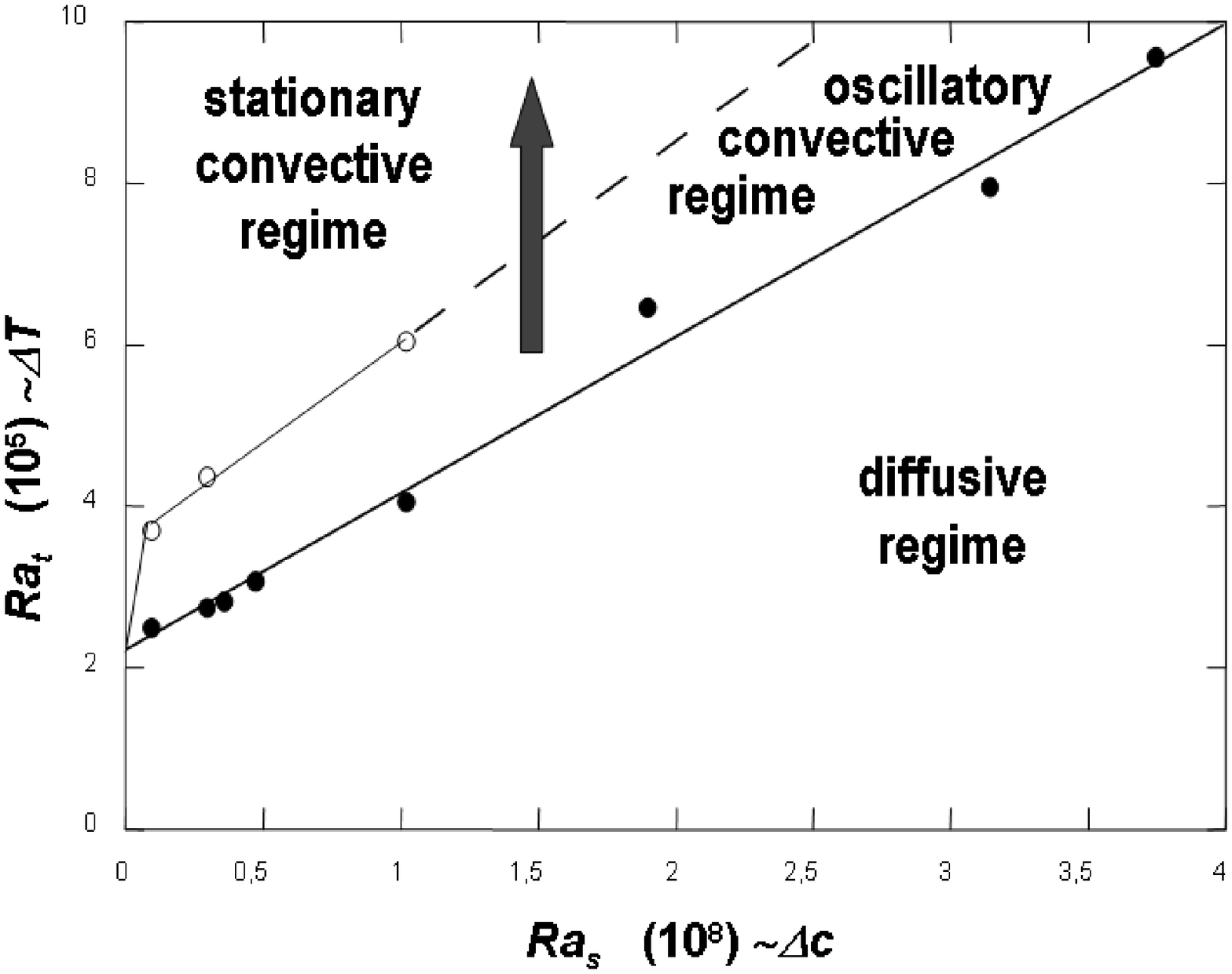}
\caption[Hydrodynamic stability diagram of (LiCl,H$_2$O) in the thermal Rayleigh
number (dimensionless number proportional to the temperature gradient) /
solutal Rayleigh number (dimensionless number proportional to the concentration
gradient) plane.
The arrow shows the evolution in the plane during the experiment of Figure
\ref{audela}.]{}
\label{diagstab}
\end{center}
\end{figure}

\begin{figure}
\begin{center}
\includegraphics[width=.6\textwidth]{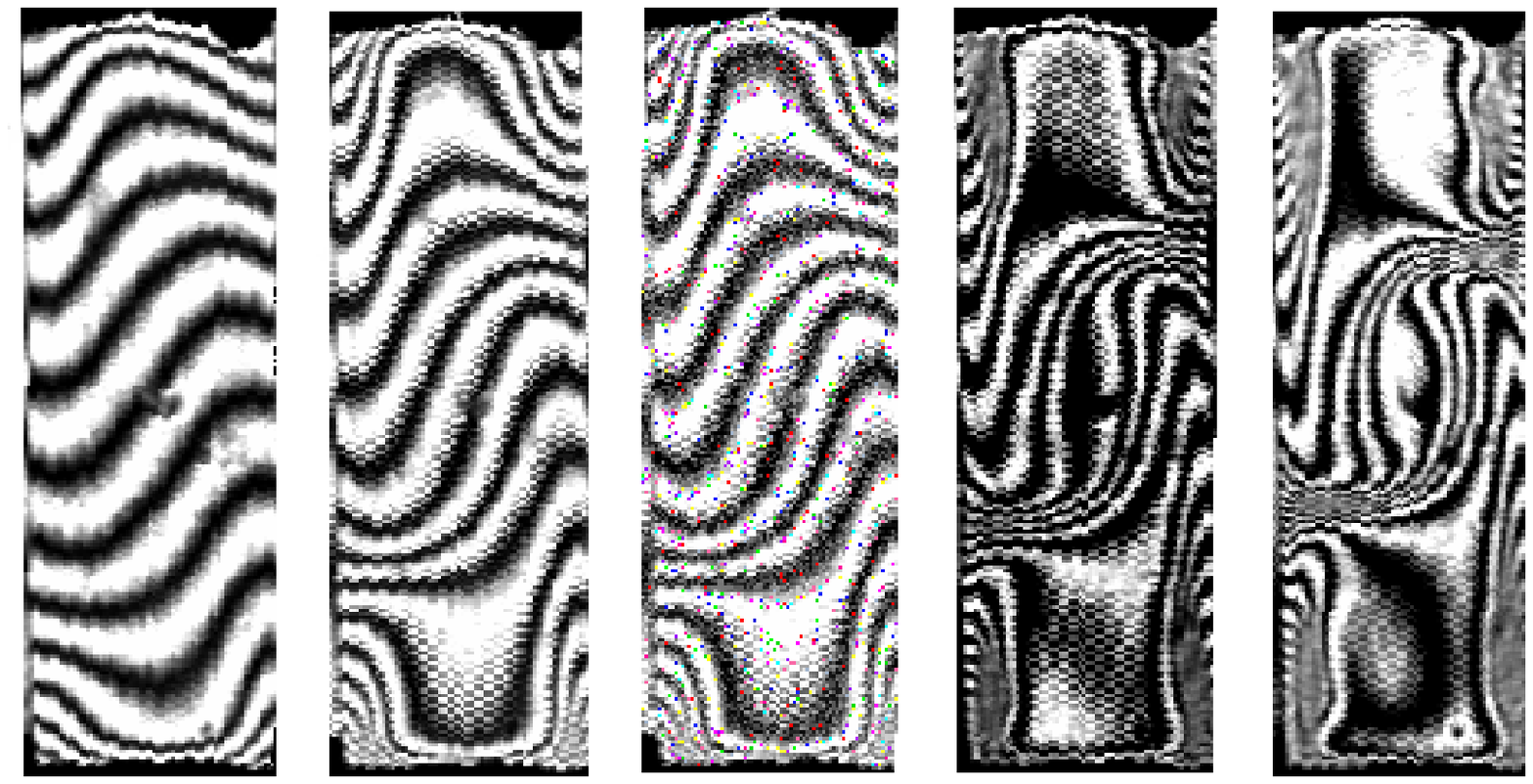}
\caption[Temperature fringes in (LiCl,H$_2$0) in the stationary convective
regime for 10.5, 15.3, 18.7, 28.2 and 34.3$^{\circ}$C
temperature difference between top and bottom.]{}
\label{audela}
\end{center}
\end{figure}

Thereby, the stability diagram of the liquid has been determined: depending
on the thermal and solutal gradients, the liquid shows a diffusive, oscillatory
convective (roll changing periodically of rotation direction) or stationary
convective regime (cf. Figure \ref{diagstab}) \cite{Colombani98,Colombani99}.
The oscillatory behaviour is a consequence of the competing influence of the
stabilizing (concentration) and unstabilizing (temperature) fields.
In the stationary convective regime,
for increasing temperature gradients, other rolls appear
in corners of the cell, leading to complex convective patterns (cf. Figure
\ref{audela}).

\subsection{Dissolution}

\begin{figure}
\begin{center}
\includegraphics[width=.3\linewidth,angle=90]{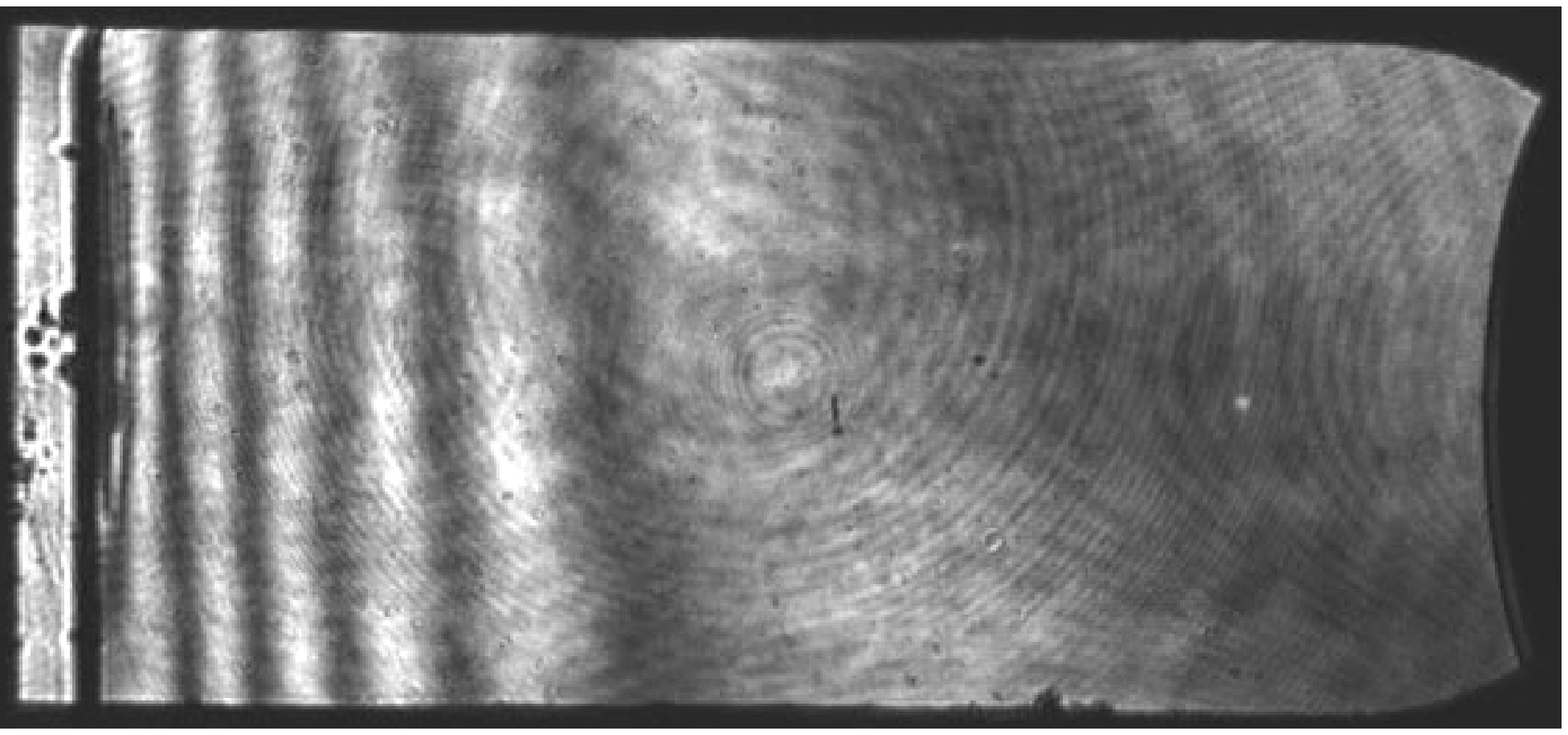}
\caption[Interference pattern due to the diffusion of dissolved matter
during the dissolution of a gypsum (CaSO$_4$,2H$_2$O)
monocrystal in pure water 240 min. after the beginning
of the experiment.]{}
\label{interf}
\end{center}
\end{figure}

\begin{figure}
\begin{center}
\includegraphics[width=.5\linewidth,angle=-90]{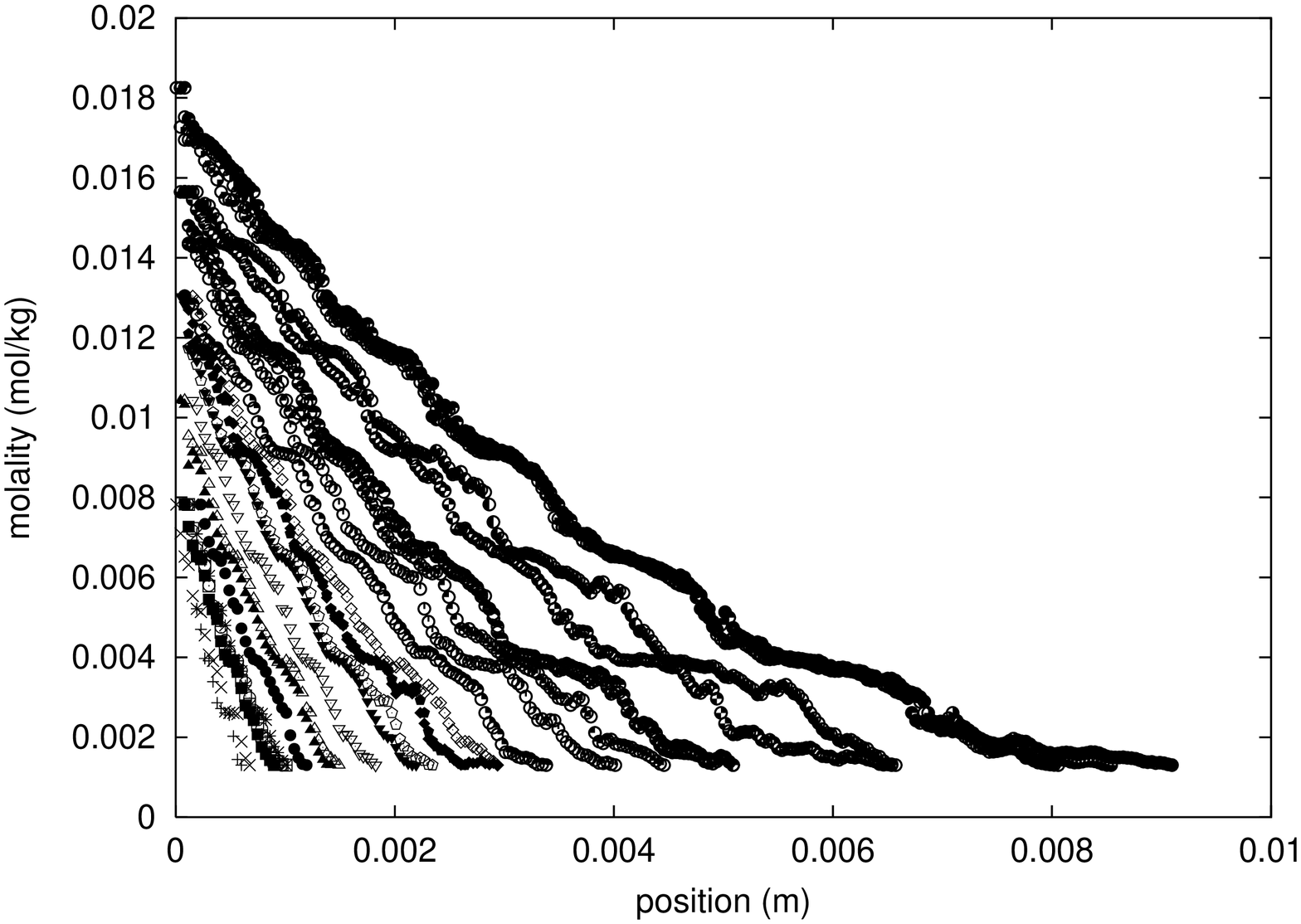}
\caption[Molality evolution in the cell for times ranging
from 0 to 330 min. in the experiment of Figure \ref{interf}.]{}
\label{molality}
\end{center}
\end{figure}

We are now going to consider the possible interest of this technique for the
study of chemical reactions.
Since the beginning of the technique \cite{Knox} and up to now \cite{Arise},
solution chemistry, electrochemistry, corrosion study or crystal growth
investigation appeal to holographic interferometry potentialities to
measure quantitative transfer rates at solid-liquid interfaces.
For instance the study of dissolution is particularly appropriate.
Holographic interferometry enables indeed a simultaneous measurement of the
dissolution coefficient, diffusion coefficient and solubility of the material.
During an experiment, the reference hologram of the pure liquid is recorded
and a sample of the dissolving solid is subsequently introduced in the cell.
The solid reacts at the solid-liquid interface and then diffuses in the liquid.
The resulting concentration change in the liquid is visualized via
concentration fringes (cf. Figure \ref{interf}).
Using again the expression of the molality evolution between contiguous
fringes $\delta m=\lambda(e(\partial n/\partial m))$,
the molality profile in the whole cell can be plotted, as shown in
Figure \ref{molality}.

To evaluate the analytical expression of the substance molality $m$
(or other concentration unit) in the liquid, Fick's equation
$\left(\frac{\partial m}{\partial t}\right)_z=
D\left(\frac{\partial^2 m}{\partial z^2}\right)_t$
must be solved with relevant boundary conditions.
At the dissolving interface, the boundary condition equalizes the dissolution
flux and the diffusion flux.
In order to obtain an analytical mathematical expression,
the liquid-air meniscus is considered as being infinitely distant from the
reaction.
The resulting integration leads to :
\begin{equation}
m(z,t)=m^{sat}\left[\text{erfc}\left(\frac{z}{2\sqrt{Dt}}\right)-\exp(hz+h^2Dt)
\text{erfc}\left[\frac{z}{2\sqrt{Dt}}+h\sqrt{Dt}\right]\right].
\label{semiinf}
\end{equation}
In this equation, the parameter
$h=k\beta/(D\rho m^{sat})$ has been introduced with
$k$ the dissolution coefficient, $\beta$ a geometrical parameter, $D$ the
diffusion coefficient, $\rho$ the density, $m^{sat}$ the solubility and $L$
the total diffusing length (height of the experimental cell).
The best fit of the experimental molality profiles with this theoretical
expression brings the expected $m^{sat}$, $D$ and $k$ values
\cite{Colombani06}.

\section{Conclusion}

Holographic interferometry makes use of the capacity of wavefront
reconstruction of holography to perform
interferences between the same phase object at different times.
This possibility gives
access to the real time evolution of the 2D field (integrated along the
optical path) of refraction index in liquids.
The cartography of all parameters influencing the refraction index is thus
accessible, temperature and concentration essentially.
The possible use of this investigation method of liquids are mainly:
measurement of kinetic coefficients (mass diffusion, thermal diffusion,
dissolution) and visualization of flows.
The continuous progress in digital holographic interferometry still expands the
possible contexts of utilisation of this technique.

\section*{Acknowledments}
We acknowledge CNES (french space agency) for financial support.


\newpage

\begin{thebibliography}{10}

\bibitem{Gabor48}
D.~Gabor,  Nature 161 (1948) 777.

\bibitem{Leith}
E.~Leith and J.~Upatnieks,  J. Opt. Soc. Am. 52 (1962) 1123.

\bibitem{Vest}
C.~Vest, {\em Holographic interferometry}, Wiley, New York, 1979.

\bibitem{Horman}
M.~Horman,  Appl. Opt. 4 (1965) 333.

\bibitem{Gabor65}
D.~Gabor, G.~Stroke, R.~Restrick, A.~Funkhouser, and D.~Brumm,  Phys.
  Lett. 18 (1965) 116.

\bibitem{Brooks}
R.~Brooks, L.~Heflinger, and R.~Wuerker,  Appl. Phys. Lett.
  7 (1965) 248.

\bibitem{Goodman}
J.~Goodman and R.~Lawrence,  Appl. Phys. Lett. 11 (1967) 77.

\bibitem{Becsey71}
J.~Becsey, N.~Jackson, and J.~Bierlein,  J. Phys. Chem. 75 (1971)
  3374.

\bibitem{Budziak}
A.~Budziak, M.~Zimnal, and A.~Czapkiewicz,  Acta Phys. Polon.
  A40 (1971) 545.

\bibitem{Durou72}
C.~Durou, J.~Giraudou, and J.~Mahenc,  C.R. Acad. Sc. Paris S\'erie C
 275 (1972) 761.

\bibitem{Durou73}
C.~Durou,  J. Phys. E: Sci. Instr. 6 (1973) 1116.

\bibitem{Bochner}
N.~Bochner and J.~Pipman,  J. Phys. D: Appl. Phys. 9 (1976) 1825.

\bibitem{Anand}
A.~Anand, V.~Chhaniwal, and C.~Narayanamurthy,  Appl. Opt.
  45 (2006) 904.

\bibitem{Baranski}
J.~Baranski, E.~Bich, E.~Vogel, and J.~Lehmann,  Int. J. Thermophys.
 24 (2003) 1207.

\bibitem{Roetzel}
W.~Roetzel, D.~Bl\"omker, and W.~Czarnetzki,  Chem. Ing. Tech.
 69 (1997) 674.

\bibitem{Reyes}
L.~Reyes, J.~Bert, J.~Fornazero, R.~Cohen, and L.~Heinrich,  Coll. Surf. B:
  Biointerf. 25 (2002) 99.

\bibitem{Aouizerat}
A.~Aouizerat-Elarby, H.~Dez, B.~Prevel, J.~Jal, J.~Bert, and J.~Dupuy-Philon,
   J. Mol. Liq. 84 (2000) 289.

\bibitem{Colombani01}
J.~Colombani, B.~Herv\'e, and J.~Bert,  J. Phys. IV (France)
  11 Pr6 (2001) 35.

\bibitem{Soret}
C.~Soret,  Arch. Sci. Phys. Nat. Gen\`eve 2 (1879) 48.

\bibitem{Becsey70}
J.~Becsey, G.~Maddux, N.~Jackson, and J.~Bierlein,  J. Phys. Chem.
 74 (1970) 1401.

\bibitem{Sanchez73}
V.~Sanchez, C.~Durou, and J.~Mahenc,  C.R. Acad. Sci. C 277 (1973)
  663.

\bibitem{Sanchez76}
V.~Sanchez and J.~Mahenc,  J. Chim. Phys. (Paris) 73 (1976) 485.

\bibitem{Colombani98}
J.~Colombani, H.~Dez, J.~Bert, and J.~Dupuy-Philon,  Phys. Rev. E
 58 (1998) 3202.

\bibitem{Colombani99}
J.~Colombani, J.~Bert, and J.~Dupuy-Philon,  J. Chem. Phys.
 110 (1999) 8622.

\bibitem{Knox}
C.~Knox, R.~Sayano, E.~Seo, and H.~Silverman,  J. Phys. Chem.
 71 (1967) 3102.

\bibitem{Arise}
I.~Arise, Y.~Fukunaka, and F.~McLarnon,  J. Electrochem. Soc.
  153 (2006) A69.

\bibitem{Colombani06}
J.~Colombani and J.~Bert, {\em Holographic interferometry study of the
  dissolution and diffusion of gypsum in water}.
\newblock submitted.

\end{thebibliography}

\clearpage

\listoffigures

\clearpage

\begin{figure}
\begin{center}
\includegraphics[width=.6\textwidth]{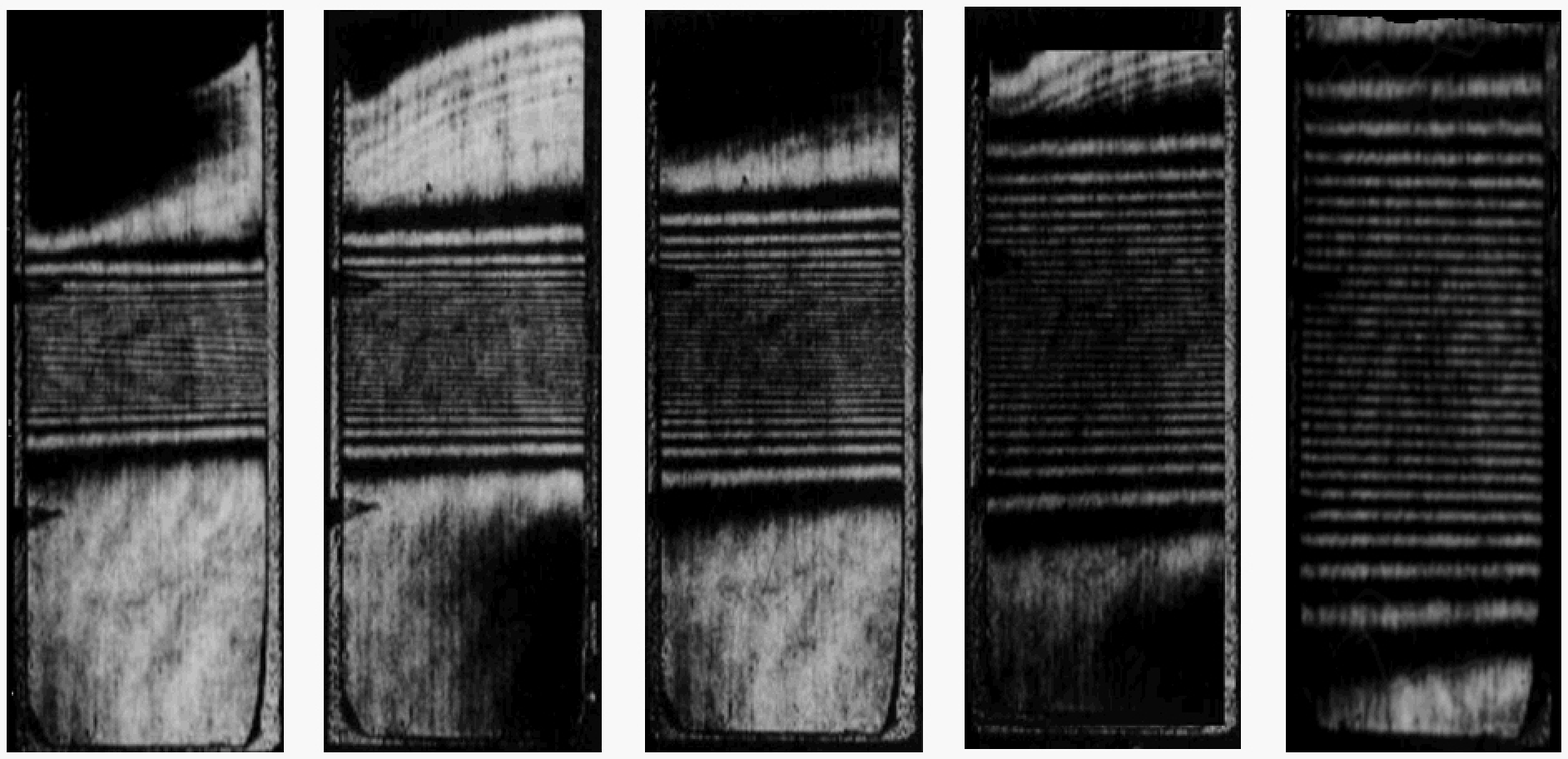}
\caption[Concentration fringes during the interdiffusion process of two
slightly different concentrations of an aqueous bovine serumalbumin solution 15,
32, 53, 62 and 71 min. after the beginning of the experiment.]{}
\label{difprot}
\end{center}
\end{figure}

\end{document}